\documentclass[12pt]{article}

\usepackage{amsmath,amssymb,amsfonts,graphics,graphicx,amscd,amsfonts,epsfig,color}
\usepackage{subfig}
\usepackage{cite}
\usepackage{enumitem}

\newcommand {\beq} {\begin{equation}}
\newcommand {\eeq} {\end{equation}}
\newcommand {\beqa}{\begin{eqnarray}}
\newcommand {\eeqa}{\end{eqnarray}}

\makeatletter
    
    \@addtoreset{equation}{section}
  \makeatother

\allowdisplaybreaks[4]

\date{}
\begin{document}

\begin{flushright} 
KEK-TH-1967
\end{flushright} 

\vspace{0.1cm}

\begin{center}
{\LARGE Combining the complex Langevin method and \\
the generalized Lefschetz-thimble method}
\end{center}
\vspace{0.1cm}
\vspace{0.1cm}
\begin{center}

         Jun N{\sc ishimura}$^{ab}$\footnote
          {
 E-mail address : jnishi@post.kek.jp} 
and
         Shinji S{\sc himasaki}$^{ac}$\footnote
          {
 E-mail address : shinji.shimasaki@keio.jp} 

\vspace{0.5cm}

$^a${\it KEK Theory Center, 
High Energy Accelerator Research Organization,\\
1-1 Oho, Tsukuba, Ibaraki 305-0801, Japan}

$^b${\it Graduate University for Advanced Studies (SOKENDAI),\\
1-1 Oho, Tsukuba, Ibaraki 305-0801, Japan} 

$^c${\it Research and Education Center for Natural Sciences, Keio University,\\
Hiyoshi 4-1-1, Yokohama, Kanagawa 223-8521, Japan}

\end{center}

\vspace{1.5cm}

\begin{center}
  {\bf abstract}
\end{center}

\noindent 
The complex Langevin method and 
the generalized Lefschetz-thimble method 
are two closely related approaches 
to the sign problem,
which are both 
based on complexification of the original dynamical variables.
The former can be viewed as a generalization of 
the stochastic quantization using the Langevin equation,
whereas the latter is
a deformation of the integration contour using the so-called
holomorphic gradient flow.
In order to clarify their relationship,
we propose a formulation which combines
the two methods by applying the former method
to the real variables that parametrize
the deformed integration contour in the latter method.
Three versions, which differ in the treatment
of the residual sign problem in the latter method, are considered.
%
By applying them to a single-variable model,
we find, in particular, that one of the versions interpolates 
the complex Langevin method and the original Lefschetz-thimble method.

\newpage

\section{Introduction}

The sign problem is a notorious problem in computational physics,
which occurs in performing a multi-variable integral with a complex weight
based on importance sampling such as Monte Carlo simulation.
Recently two closely related
methods have been pursuit as promising solutions
to this problem
based on complexification of the original real variables.
One is the complex Langevin method (CLM) \cite{Parisi:1984cs,Klauder:1983sp},
and the other is the Lefschetz-thimble 
method \cite{Cristoforetti:2012su}. 
An important generalization of the second approach was 
proposed \cite{Alexandru:2015sua}
to overcome some problems in the original proposal,
and we will therefore refer to this approach as
the generalized Lefschetz-thimble method (GLTM) in this paper.

The CLM is an extension of the 
stochastic quantization \cite{Parisi:1980ys} 
(See ref.~\cite{Damgaard:1987rr} for a review.),
which generates dynamical variables
with a given probability 
by solving the Langevin equation
that describes a fictitious time evolution of those variables
under the influence of a Gaussian noise.
In the case of a complex weight,
one has to complexify the dynamical variables,
and the drift term in the Langevin equation
should be extended to a holomorphic function of the complexified variables
by analytic continuation. 
When one measures
the observables for the complexified variables
generated by the Langevin process,
one also has to extend the observables
to holomorphic functions of the complexified variables.
Then, under certain conditions, one can 
show \cite{Aarts:2009uq,Aarts:2011ax}
that the expectation values of the observables calculated this way
at some Langevin time 
are equal to 
the expectation values of the observables 
for the original real variables with a complex weight,
which evolves with the Langevin time 
following the Fokker-Planck equation.
If this is the case, 
one can obtain the desired expectation values 
in the long Langevin-time limit \cite{Nishimura:2015pba}.
Recently, a subtlety in the use of time-evolved observables
in the original argument \cite{Aarts:2009uq,Aarts:2011ax}
was recognized \cite{Nagata:2016vkn}.
This subtlety was fixed in a refined argument, 
which also led to the derivation of a practical criterion 
for correct convergence in the CLM \cite{Nagata:2016vkn}.

On the other hand, the GLTM \cite{Alexandru:2015sua} is based 
on a continuous deformation of the integration contour 
using the so-called holomorphic gradient flow.
Along the deformed contour, 
the phase of the complex weight fluctuates only mildly 
for a sufficiently long flow time, and 
one can apply ordinary Monte Carlo methods using
the absolute value of the weight in generating configurations and 
reweighting the phase factor when one calculates the expectation values.
In the long flow-time limit, the deformed contour
becomes a set of Lefschetz thimbles
and the GLTM reduces to the 
originally proposed method \cite{Cristoforetti:2012su}.
(See refs.~\cite{Cristoforetti:2013wha,Mukherjee:2014hsa,%
Fujii:2013sra,DiRenzo:2015foa,%
Tanizaki:2015rda,Fujii:2015vha,Alexandru:2015xva,%
Fukushima:2015qza,%
Tanizaki:2016cou,%
Tanizaki:2016xcu,%
Fukuma:2017fjq, Alexandru:2017oyw}
for related work.)
The phase of the complex weight becomes constant
on each thimble, which makes this limit attractable at first sight.
However, when there exists more than one thimbles,
the transition from one to another does not occur
during the Monte Carlo simulation and hence the
ergodicity is violated in this limit.
Therefore, there is an optimal flow time 
in general \cite{Alexandru:2016ejd,Alexandru:2016san,Alexandru:2016gsd}.
(See ref.~\cite{Fukuma:2017fjq, Alexandru:2017oyw} for a new proposal concerning this issue.)

Let us recall here that the two methods have their own pros and cons.
From the viewpoint of solving the sign problem,
the CLM is more powerful since it solves it completely as far as
it works, whereas the GLTM has a residual sign problem, which
may be problematic when the system size becomes large.
From the viewpoint of numerical cost, 
the GLTM is more demanding since one has to calculate the
Jacobian arising from deforming the integration contour.
(See, however, ref.~\cite{Alexandru:2016lsn}.)
It should also be noted that
one can include fermions in the CLM
with the cost of O($V$) for the system size $V$, whereas
in the GLTM, it requires O($V^3$).
The disadvantage of 
the CLM is that 
there is a 
condition \cite{Aarts:2009uq,Aarts:2011ax,Nagata:2016vkn}
to be satisfied to make it work, and hence there is a certain
range of applicability.
While the gauge cooling technique \cite{Seiler:2012wz} 
(See ref.~\cite{Nagata:2016vkn,Nagata:2015uga} for its justification)
has enlarged this range of applicability to the extent that
finite density QCD either
with heavy quarks \cite{Aarts:2014bwa,Aarts:2013uxa,Aarts:2016qrv}
or in the deconfined phase \cite{Sexty:2013ica,Fodor:2015doa} can 
now be investigated, 
it is not yet clear whether one can investigate it 
even in the confined phase 
with light quarks \cite{Mollgaard:2013qra,Mollgaard:2014mga,%
Sinclair:2016nbg,%
Nagata:2016alq,Ito:2016efb,Salcedo:2016kyy,Aarts:2017vrv,Bloch:2016jwt}.

In this situation, it is useful to compare the two methods
closely \cite{Aarts:2013fpa,Aarts:2014nxa}
and to try to come up with a new method which is free from
the aforementioned problems of each 
method \cite{Tsutsui:2015tua,Hayata:2015lzj}.
Note here that despite the resemblance of the two methods in that they 
both use complexified variables, 
the connection between them is far from obvious.
In particular, the CLM samples configurations in the 
complex plane based on the Langevin process,
whereas the GLTM samples configurations on a particular contour
in the complex plane
obtained by deforming the original contour along the real axis.
Therefore the dimension of the configuration space 
in the CLM is twice as large as that of the GLTM.

In order to clarify the relationship between the two methods,
we propose a formulation which combines
them
by applying the CLM to the real variables
that parametrize the deformed integration contour in the GLTM.
We consider three versions, which differ in the treatment
of the residual sign problem in the GLTM.
The ordinary CLM and the GLTM are included 
in our combined formulation as special cases.
We apply our formulation to
a simple one-variable case with a single Lefschetz thimble,
and investigate the distribution of flowed configurations in detail.
We find, in particular, that one version of our formulation
interpolates the ordinary CLM and the original Lefschetz-thimble method.

The rest of this paper is organized as follows.
In section \ref{sec:review}
we briefly review the CLM and the GLTM.
In section \ref{sec:unify} we present our formulation,
which combines the two methods.
In section \ref{sec:0d-model}
we apply our formulation to
a single-variable case and
clarify the relationship between the two methods.
Section \ref{sec:summary}
is devoted to a summary and discussions.

\section{Brief review of the CLM and the GLTM}
\label{sec:review}

In this section, we briefly review the CLM and the GLTM.
Here we consider a general model given by a multi-variable integral 
\begin{align}
Z=\int_{\mathbb{R}^n} dx \, e^{-S(x)} \ ,
\label{Z}
\end{align}
where the action $S(x)$ is a complex function of 
$x=(x_1,\cdots,x_n) \in \mathbb{R}^n$.
The expectation value of an observable $\mathcal O(x)$ is defined by
\begin{align}
\langle \mathcal O(x)\rangle =
\frac{1}{Z}\int_{\mathbb{R}^n} dx \, \mathcal O(x) \, e^{-S(x)} \ .
\label{O}
\end{align}

\subsection{Complex Langevin Method (CLM)}

In the CLM, we complexify the dynamical variables 
$x\in\mathbb{R}^n$ as $x\to z= x+iy\in \mathbb{C}^n$
and extend the action and the observable to holomorphic functions of $z$.
Then we consider the complex Langevin equation
\begin{align}
\frac{\partial}{\partial t}z_k(t)
&=-\frac{\partial S(z)}{\partial z_k}+\eta_k(t) \ ,
\label{cle}
\end{align}
where $t$ is a fictitious time and 
$\eta_k(t)$ is a real Gaussian noise normalized as 
$\langle \eta_k(t)\eta_l(t') \rangle_\eta = 2 \, \delta_{kl} \, \delta(t-t')$.
This equation (\ref{cle}) defines a probability distribution of 
$z$ at each time $t$.
It is shown \cite{Aarts:2009uq,Aarts:2011ax} that under certain conditions,
the expectation value $\langle\mathcal O(x)\rangle$
in the original integral \eqref{O}
can be obtained as
\begin{align}
\langle\mathcal O(x)\rangle = \langle \mathcal O(z)\rangle_{\mathrm{CLM}} \ ,
\label{clm ob}
\end{align}
where the right-hand side
denotes the expectation value of $\mathcal O(z)$ 
with respect to the limiting distribution of $z$ at $t\to \infty$.
As in the case of Monte Carlo methods,
$\langle \mathcal O(z)\rangle_{\mathrm{CLM}}$
can be replaced with a long-time average of the observable
calculated for the generated configurations $z(t)$
if ergodicity holds for the Langevin time-evolution.

The derivation of the equality \eqref{clm ob}
uses integration by parts, which can be justified
if the distribution of $z$ falls off fast enough
in the imaginary directions \cite{Aarts:2009uq,Aarts:2011ax}
as well as near the singularities of the drift term 
if they exist \cite{Nishimura:2015pba,Aarts:2017vrv}.
Recently \cite{Nagata:2016vkn}, a subtlety in the use of time-evolved 
observables in the original argument \cite{Aarts:2009uq,Aarts:2011ax} 
was pointed out, and the derivation of the equality \eqref{clm ob}
has been refined taking account of this subtlety.
This also led to the proposal of a useful criterion for justification,
which states that
the distribution of the drift term should 
be suppressed exponentially or faster at large magnitude \cite{Nagata:2016vkn}.

\subsection{Generalized Lefschetz-Thimble Method (GLTM)}

In the GLTM \cite{Alexandru:2015sua}, 
we deform the integration contour from $\mathbb{R}^n$ 
to an $n$-dimensional real manifold in $\mathbb{C}^n$
by using the so-called holomorphic gradient flow,
which makes the sign problem milder.

The holomorphic gradient flow is defined by
the differential equation 
%
\begin{align}
\frac{\partial}{\partial \sigma}\phi_k(x;\sigma) 
= \overline{ \left(\frac{\partial S(\phi(x;\sigma))}{\partial \phi_k}\right)} \ ,
\label{hge}
\end{align}
which is solved from $\sigma = 0$ to $\sigma=\tau$
with the initial condition $\phi(x;0)=x\in \mathbb{R}^n$.
The flowed configurations define
a $n$-dimensional real manifold in $\mathbb{C}^n$,
which we denote as $M_\tau=\{\phi(x;\tau)|x\in\mathbb{R}^n\}$.
One can actually argue that
the integration contour can be deformed continuously
from $\mathbb{R}^n$ to $M_\tau\subset \mathbb{C}^n$
without changing the partition function.
Then, by noting that 
$\phi(x)\equiv\phi(x;\tau)$
defines a one-to-one map
from $x\in \mathbb{R}^n$ to $\phi\in M_\tau$,
one can rewrite the partition function as
\begin{align}
Z
=\int_{M_\tau} d\phi \, e^{-S(\phi)}
=\int_{\mathbb{R}^n} dx \, \mathrm{det}J(x) \, e^{-S(\phi(x))} \ ,
\label{gltmZ}
\end{align}
where $J_{kl}(x)$
is the Jacobian corresponding to the map $\phi(x)$.
We obtain the Jacobian as $J_{kl}(x)= J_{kl}(x;\tau)$, where
$J_{kl}(x;\sigma) \equiv \partial \phi_k(x;\sigma)/\partial x_l$
is calculated by solving the differential equation
\begin{align}
\frac{\partial}{\partial \sigma}J_{kl}(x;\sigma)
=\overline{\left(\frac{\partial^2 S(\phi(x;\sigma))}{\partial \phi_k\partial \phi_m} 
J_{ml}(x;\sigma)\right)} \ ,
\label{hgeJ}
\end{align}
from $\sigma = 0$ to $\sigma=\tau$
with the initial condition $J_{kl}(x;0)=\delta_{kl}$. 
Thus, the deformation of the integration contour
simply amounts to changing the action from $S(x)$ to the ``effective action''
\begin{align}
S_{\mathrm{eff}}(x)=S(\phi(x)) - \log \mathrm{det} J(x) \ .
\label{Seff}
\end{align}

The virtue of using the holomorphic gradient flow
\eqref{hge} in deforming the integration contour
lies in the fact that
the real part of the action $S$ grows monotonically along the flow
keeping the imaginary part constant.
Thus, for a sufficiently large flow time $\tau$,
the partition function \eqref{gltmZ} is dominated by a small region of $x$,
and the sign problem becomes much milder.
One can therefore apply a standard Monte Carlo method
using only the real part of the effective action \eqref{Seff},
dealing with the imaginary part
by reweighting.
Thus, the expectation value of $\mathcal O(x)$ can be calculated as
\begin{align}
\langle \mathcal O(x)\rangle
=\frac{\langle e^{-i\mathrm{Im}S_{\mathrm{eff}}(x)}
\mathcal O(\phi(x;\tau))\rangle_{\mathrm{MC}}}
{\langle e^{-i\mathrm{Im}S_{\mathrm{eff}}(x)}\rangle_{\mathrm{MC}}} \ ,
\label{gltm ob}
\end{align}
where $\langle\cdots\rangle_{\mathrm{MC}}$ represents 
the expectation value obtained by a Monte Carlo method
with the weight $\exp(-\mathrm{Re}S_{\mathrm{eff}}(x))$.

When the flow time $\tau$ becomes infinitely large, $M_\tau$ contracts 
to a set of Lefschetz thimbles,
and the GLTM reduces to the so-called 
Lefschetz-thimble method \cite{Cristoforetti:2012su}.
Since $\mathrm{Im}S(\phi(x))$ is constant 
on each thimble, the sign problem is maximally reduced in some sense.
However, 
when there are more than one thimbles,
the transition between thimbles does not occur 
during the Monte Carlo simulation,
which leads to the violation of 
ergodicity.
The GLTM \cite{Alexandru:2015sua}
avoids this problem of the original method
by choosing a large but finite flow time.
Recently, it has been pointed out \cite{Fukuma:2017fjq, Alexandru:2017oyw} that
the flow time can be made as large as one wishes
without the ergodicity problem 
if one uses the parallel tempering algorithm.

\section{Combining the CLM and the GLTM}
\label{sec:unify}

In this section, we present our formulation, which combines
the CLM and the GLTM.
In fact, we consider three versions of our formulation, which differ
in the treatment of the residual sign problem in the GLTM.
The simplest version 
removes the residual sign problem completely
by applying the CLM to the partition function \eqref{gltmZ}
that appears in the GLTM.
The real variables $x\in \mathbb{R}^n$ 
parametrizing the $n$-dimensional real manifold in $\mathbb{C}^n$
have to be complexified, and they evolve in time following the
complex Langevin equation obtained from the effective action (\ref{Seff}).
The observable should be evaluated with 
the flowed configuration $\phi(z)$, which is defined 
as holomorphic extension of $\phi(x)$.
Therefore, it is important to see
how the distribution of flowed configurations
changes as the flow time increases.




In order to apply the CLM to the partition function \eqref{gltmZ},
we consider
the drift term
\begin{align}
\frac{\partial S_{\mathrm{eff}}(x)}{\partial x_k}=
\frac{\partial S(\phi(x))}{\partial \phi_l}J_{lk}(x)
- J^{-1}_{lm}(x)K_{mlk}(x) \ ,
\label{comb drift x}
\end{align}
where we have defined 
$K_{mlk}(x)\equiv \partial J_{ml}(x;\tau)/\partial x_k$,
which can be obtained by solving the differential equation
\begin{align}
\frac{\partial}{\partial \sigma}K_{klm}(x;\sigma)
&=\overline{\left(\frac{\partial^3 S(\phi(x;\sigma))}
{\partial\phi_k\partial\phi_p\partial\phi_q}
J_{pl}(x;\sigma)J_{qm}(x;\sigma)
+\frac{\partial^2S(\phi(x;\sigma))}{\partial \phi_k\partial\phi_p}
K_{plm}(x;\sigma)\right)} \ ,
\label{hgeK}
\end{align}
for $K_{mlk}(x;\sigma)\equiv \partial J_{ml}(x;\sigma)/\partial x_k$
from $\sigma=0$ to $\sigma=\tau$ 
with the initial condition $K_{mlk}(x;0)=0$.
Then the complex Langevin equation corresponding to \eqref{gltmZ}
reads
\begin{align}
\frac{\partial}{\partial t}z_k(t)
&=-\frac{\partial S(\phi(z))}{\partial \phi_l}J_{lk}(z)
+J^{-1}_{lm}(z)K_{mlk}(z)
+\eta_k(t) \ ,
\label{cle Seff}
\end{align}
where $\phi(z)$, $J_{kl}(z)$ and $K_{mlk}(z)$
are holomorphic extension of $\phi(x)$, $J_{kl}(x)$ and $K_{mlk}(x)$, 
respectively.

The crucial point to note here is that
the flow equations \eqref{hge}, \eqref{hgeJ} and \eqref{hgeK}
for the three functions $\phi(x)$, $J_{kl}(x)$ and $K_{mlk}(x)$
involve complex conjugation on the right-hand side.
Therefore, in order to define
$\phi(z;\sigma)$, $J_{kl}(z;\sigma)$ and $K_{mlk}(z;\sigma)$
as holomorphic functions of $z$,
we have to extend the flow equations as
\begin{align}
\frac{\partial}{\partial \sigma}\phi_k(z;\sigma) 
&= \overline{ \left(\frac{\partial S(\phi(\bar z;\sigma))}{\partial \phi_k}\right)} \ ,
\label{hol hge-phi}
\\[3mm]
\frac{\partial}{\partial \sigma}J_{kl}(z;\sigma)
&=\overline{\left(\frac{\partial^2 S(\phi(\bar z;\sigma))}
{\partial \phi_k\partial \phi_m}
J_{ml}(\bar z;\sigma)\right)} \ ,
\label{hol hge-J}
\\[3mm]
\frac{\partial}{\partial \sigma}K_{klm}(z;\sigma)
&=\overline{\left(\frac{\partial^3 S(\phi(\bar z;\sigma))}
{\partial\phi_k\partial\phi_p\partial\phi_q}
J_{pl}(\bar z;\sigma)J_{qm}(\bar z;\sigma)
+\frac{\partial^2S(\phi(\bar z;\sigma))}{\partial \phi_k\partial\phi_p}
K_{plm}(\bar z;\sigma)\right)}  \ ,
\label{hol hge}
\end{align}
where
$\phi(\bar z;\sigma)$, $J_{kl}(\bar z;\sigma)$ and $K_{mlk}(\bar z;\sigma)$
appear on the right-hand side.
These functions with $\bar z$ in the argument
obey the flow equations, which can be obtained by replacing
$z$ with $\bar z$ in the above equations.
In practice, we solve the two sets of equations simultaneously
with the initial conditions 
$\phi(z;0)=z$, $J_{kl}(z;0)=\delta_{kl}$ and $K_{klm}(z;0)=0$
for a particular value of $z$ at each Langevin step.

Using this formulation,
we can calculate
the expectation value 
\eqref{O} as
\begin{align}
\langle\mathcal O(x)\rangle 
= \langle \mathcal O(\phi(z;\tau))\rangle_{\mathrm{CLM}} \ ,
\end{align}
where the flowed configuration $\phi(z;\tau)$ has to be used
in evaluating the observable unlike in the ordinary CLM \eqref{clm ob}.
For $\tau=0$, our formulation reduces to the ordinary CLM
since $\phi(z;0)=z$, $J_{kl}(z;0)=\delta_{kl}$ and $K_{klm}(z;0)=0$.


The second version of our formulation amounts to applying the CLM
to a partially phase-quenched model
\begin{align}
Z_{\mathrm{pPQ}}
=\int_{\mathbb{R}^n} dx \, |\mathrm{det}J(x)| \, e^{-S(\phi(x))} \ ,
\label{gltmZ-p-PQ}
\end{align}
and treating the phase of $\mathrm{det}J(x)$ by reweighting.
The complex Langevin equation for \eqref{gltmZ-p-PQ} reads
\begin{align}
\frac{\partial}{\partial t}z_k(t)
&=-\frac{\partial S(\phi(z))}{\partial \phi_l}J_{lk}(z)
+\frac{1}{2}\left(J^{-1}_{lm}(z)K_{mlk}(z)
+\overline{J^{-1}_{lm}(\bar z)K_{mlk}(\bar z)}\right)
+\eta_k(t) \ ,
\label{cle |J|}
\end{align}
where $\frac{1}{2}(\cdots)$ is the holomorphic extension of 
$\partial \log |\mathrm{det}J(x)|/\partial x_k
=\mathrm{Re}(J_{lm}^{-1}(x)K_{mlk}(x))$.
We can calculate the expectation value
defined in \eqref{O} as
\begin{align}
\langle\mathcal O(x)\rangle 
= \frac{\langle \omega(z)
\mathcal O(\phi(z;\tau))\rangle_{\mathrm{CLM,~pPQ}}}
{\langle \omega(z) \rangle
_{\mathrm{CLM,~pPQ}}} \ ,
\label{comb |J| ob}
\end{align}
where $\langle\cdots\rangle_{\mathrm{CLM,~pPQ}}$ 
represents the expectation value obtained by the CLM
using \eqref{cle |J|}
and the reweighting factor $\omega(z)$ is holomorphic extension
of $\omega(x)\equiv \mathrm{det}J(x;\tau)/|\mathrm{det}J(x;\tau)|$.
In practice, we obtain $\omega(z)$ by solving the differential equation 
for $\omega(z;\sigma)$ from $\sigma=0$ to $\sigma=\tau$ with the initial
condition $\omega(z;0)=1$.
Note that $|\omega(z)| \neq 1$ in general unless $z$ is real.


The third version amounts to applying the CLM to
the (totally) phase-quenched model
\begin{align}
Z_{\rm PQ}
=\int_{\mathbb{R}^n} dx \, |\mathrm{det}J(x) \, e^{-S(\phi(x))} | \ ,
\label{gltmZ-t-PQ}
\end{align}
and treating the phase of $\mathrm{det}J(x) \, e^{-S(\phi(x))}$ 
by reweighting.
The complex Langevin equation for \eqref{gltmZ-t-PQ} reduces to the 
real one, which is given by
\begin{align}
\frac{\partial}{\partial t}x_k(t)
&=-\frac{1}{2}\left( \frac{\partial S(\phi(x))}{\partial \phi_l}J_{lk}(x)
+\overline{\frac{\partial S(\phi(x))}{\partial \phi_l}J_{lk}(x)} \right) \nonumber\\
&\qquad +\frac{1}{2}\left(J^{-1}_{lm}(x)K_{mlk}(x)
+\overline{J^{-1}_{lm}(x)K_{mlk}(x)}\right)
+\eta_k(t) \ .
\label{rle quench}
\end{align}
We can calculate the expectation value
defined in \eqref{O} as
\begin{align}
\langle\mathcal O(x)\rangle 
= \frac{\langle e^{i \Gamma(x)}
\mathcal O(\phi(x;\tau))\rangle_{\mathrm{RLM,~PQ}}}
{\langle e^{i \Gamma(x)} \rangle_{\mathrm{RLM,~PQ}}} \ ,
\label{comb PQ ob}
\end{align}
where $\langle\cdots\rangle_{\mathrm{RLM,~PQ}}$ 
represents the expectation value obtained by the real Langevin method (RLM)
using \eqref{rle quench},
and $\Gamma(x)$ in the reweighting factor $e^{i \Gamma(x)}$ is 
the phase of $\mathrm{det}J(x) \, e^{-S(\phi(x))}$.
This version is nothing but the GLTM with the RLM used 
as a standard Monte Carlo method in evaluating \eqref{gltm ob}.


\section{Results for a single-variable model}
\label{sec:0d-model}

In this section, we apply the three versions of our combined formulation
to a single-variable model to clarify
the relationship between the CLM and the GLTM.
The model is defined by the partition function
\cite{Nishimura:2015pba,Nagata:2016vkn,Aarts:2017vrv},
\begin{align}
Z=\int dx \, (x+i\alpha)^p \, e^{-x^2/2} \ ,
\label{single variable Z}
\end{align}
where $x$ is a real variable, 
while $\alpha$ and $p$ are real parameters. 
When $\alpha \neq 0$ and $p\neq 0$, the weight becomes complex and
the sign problem occurs.

First, using the holomorphic gradient flow for a fixed flow time $\tau$, 
we rewrite \eqref{single variable Z} as
\begin{align}
Z=\int dx \,  J(x)  \,  (\phi(x)+i\alpha)^p \, e^{-\phi(x)^2/2} \ ,
\label{single variable Z deform}
\end{align}
where $\phi(x)=\phi(x;\tau)$ and $J(x)=J(x;\tau)$ are 
obtained by solving \eqref{hge} and \eqref{hgeJ}
for $\phi(x;\sigma)$ and 
$J(x;\sigma) \equiv \partial \phi(x;\sigma)/\partial x$.
Then we apply the three versions of our formulation; namely
\begin{enumerate}[label=(\roman*)]
\item CLM applied to the model (\ref{single variable Z deform}).
\item CLM applied to the partially phase-quenched model:
\begin{align}
Z_{\rm pPQ}=
\int dx \,  
|J(x)| \,  (\phi(x)+i\alpha)^p \, e^{-\phi(x)^2/2} \ .
\label{single variable Z deform-pPQ}
\end{align}
\item RLM applied to the totally phase-quenched model:
\begin{align}
Z_{\rm PQ}=
\int dx \,  
|J(x) \,  (\phi(x)+i\alpha)^p  \, e^{-\phi(x)^2/2} |  \ .
\label{single variable Z deform-PQ} 
\end{align}
\end{enumerate}
In the cases (ii) and (iii), appropriate reweighting is needed
in calculating the expectation value as described in the previous section.
The phase of $J(z)$ is evaluated as the imaginary part of $\log J(z)$,
which is obtained by solving the differential equation
\begin{align}
\frac{\partial}{\partial \sigma}\log J(z;\sigma)
&=\frac{1}{J(z;\sigma)}\overline{\left(\frac{\partial^2 S(\phi(\bar z;\sigma))}
{\partial \phi^2}
J(\bar z;\sigma)\right)} 
\end{align}
from $\sigma=0$ to $\sigma=\tau$ with the initial condition $\log J(z;0)=0$.

\begin{figure}[t]
\centering
\includegraphics[width=7cm]{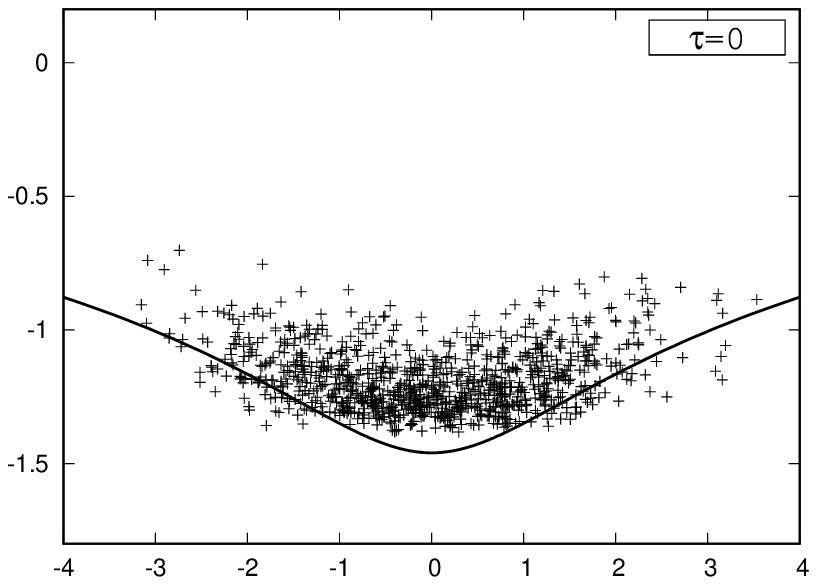}
\includegraphics[width=7cm]{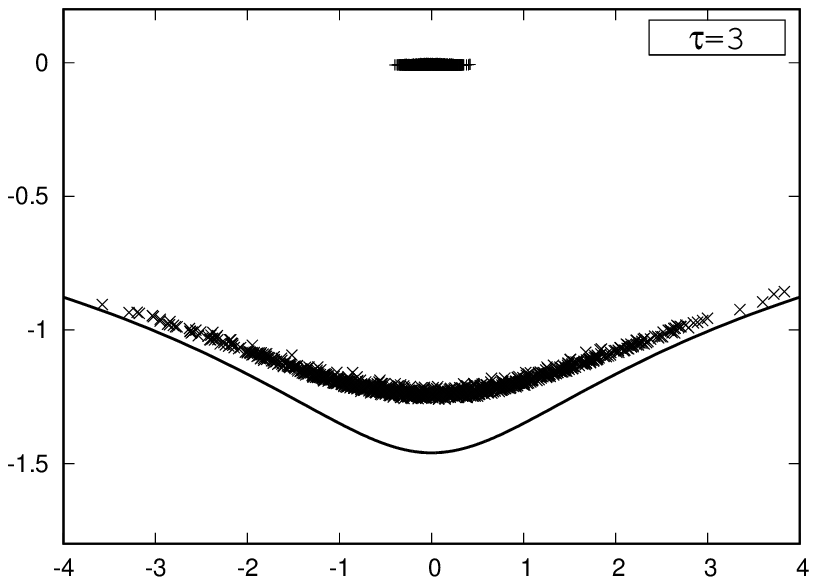}
\includegraphics[width=7cm]{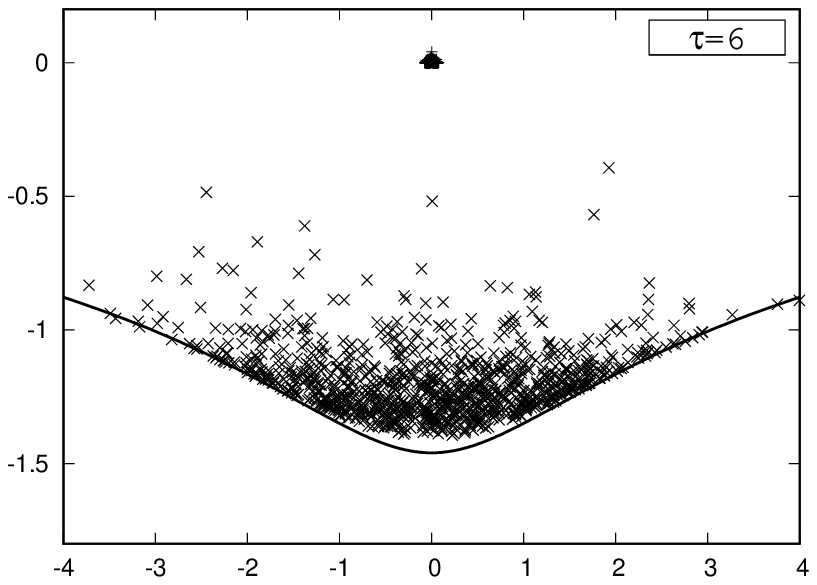}
\includegraphics[width=7cm]{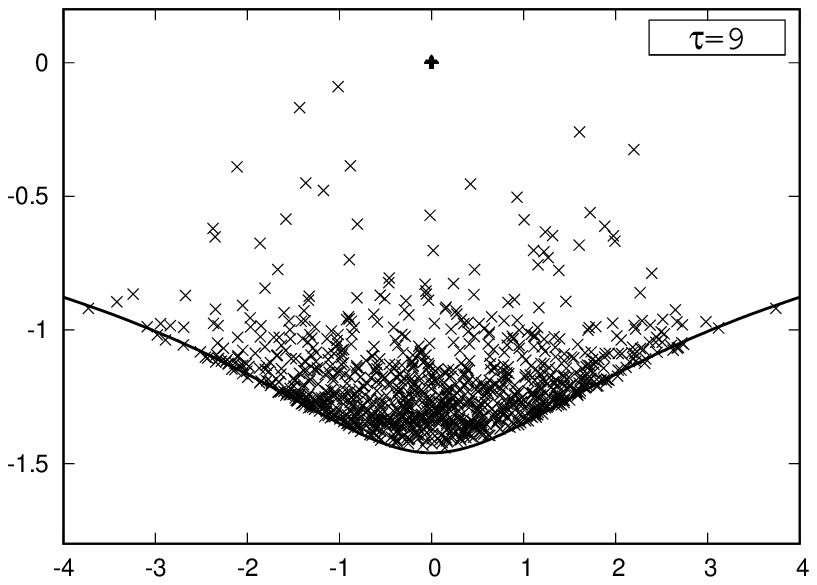}
\caption{The distribution of $z$ (represented by ``$+$'') and
flowed configurations (represented by ``$\times$'')
obtained by applying the CLM to \eqref{single variable Z deform} 
are plotted 
for $\tau=0,3,6,9$ from Top-Left to Bottom-Right.
The solid line represents the Lefschetz thimble.
At $\tau=0$, the distribution of $z$ and flowed configurations
coincides.}
\label{fig case1}
\end{figure}

\begin{figure}[t]
\centering
\includegraphics[width=7cm]{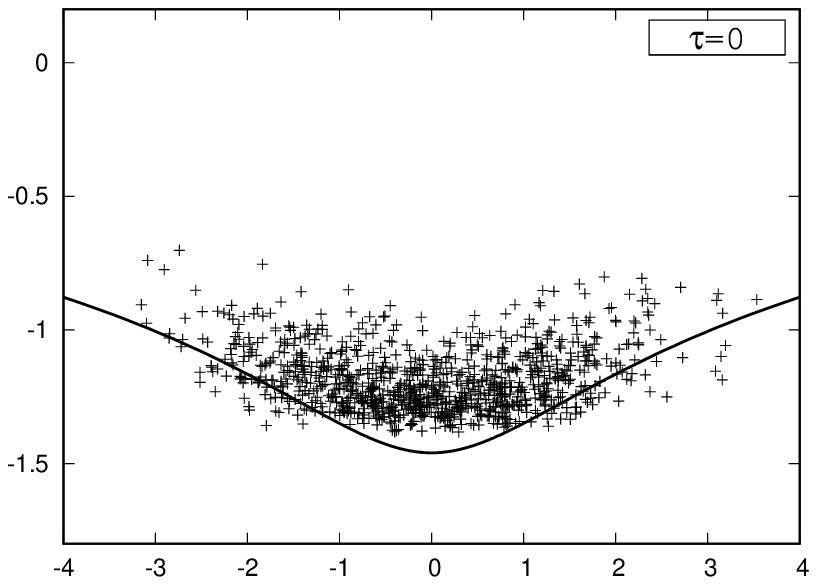}
\includegraphics[width=7cm]{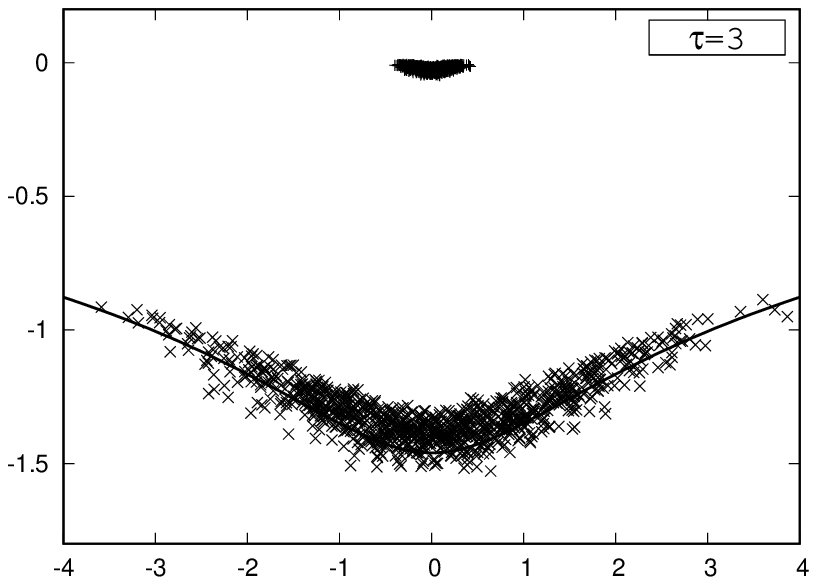}
\includegraphics[width=7cm]{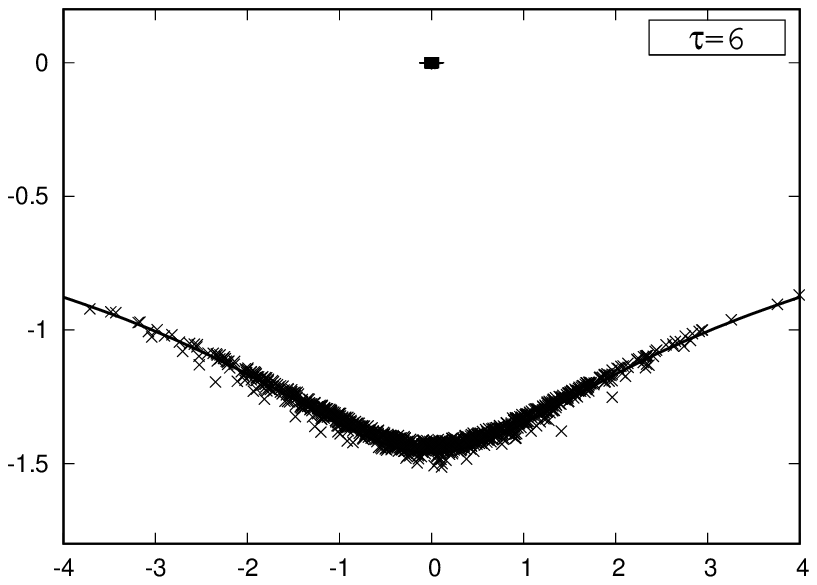}
\includegraphics[width=7cm]{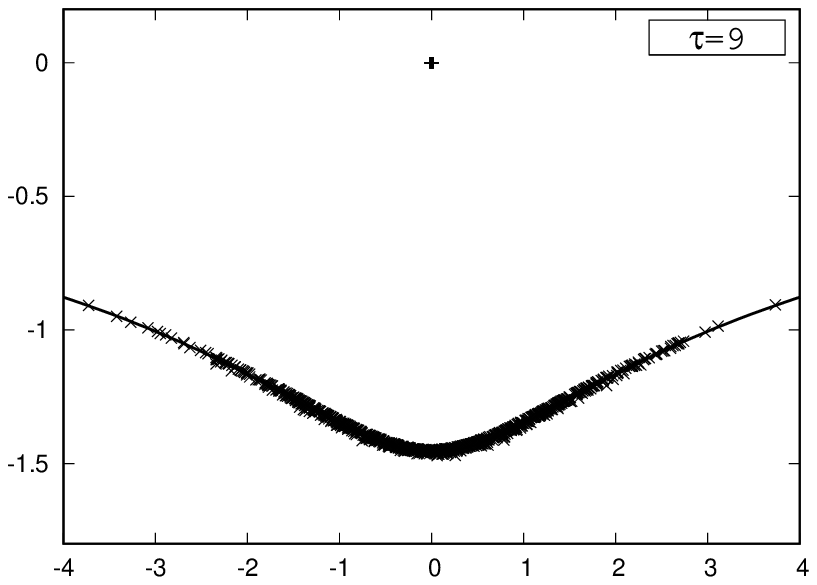}
\caption{The distribution of $z$ (represented by ``$+$'') and
flowed configurations (represented by ``$\times$'')
obtained by applying the CLM 
to the partially phase quenched model
(\ref{single variable Z deform-pPQ})
are plotted for $\tau=0,3,6,9$ from Top-Left to Bottom-Right.
The solid line represents the Lefschetz thimble.
At $\tau=0$, the distribution of $z$ and 
flowed configurations coincides.}
\label{fig case2}
\end{figure}

\begin{figure}[t]
\centering
\includegraphics[width=7cm]{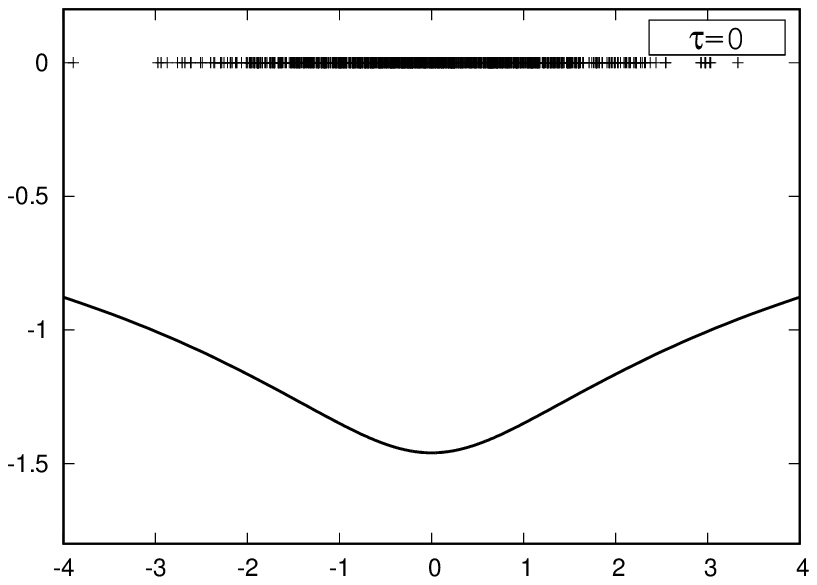}
\includegraphics[width=7cm]{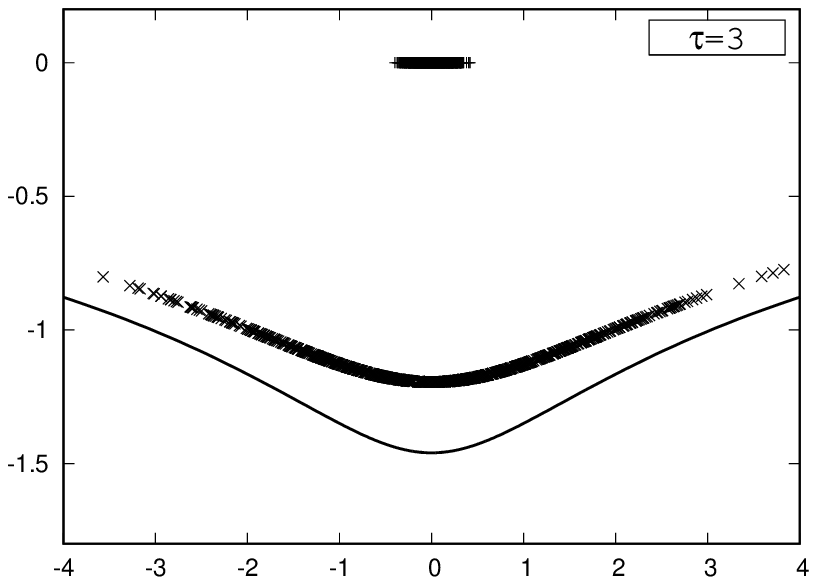}
\includegraphics[width=7cm]{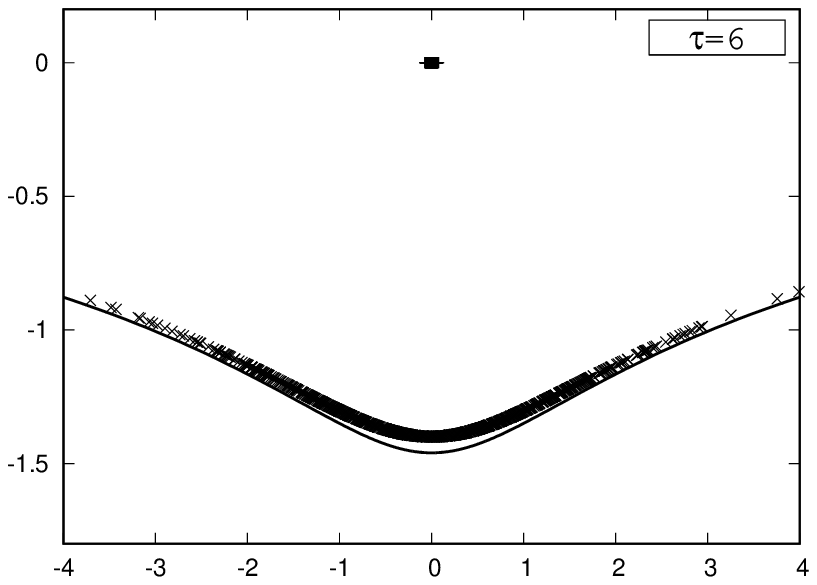}
\includegraphics[width=7cm]{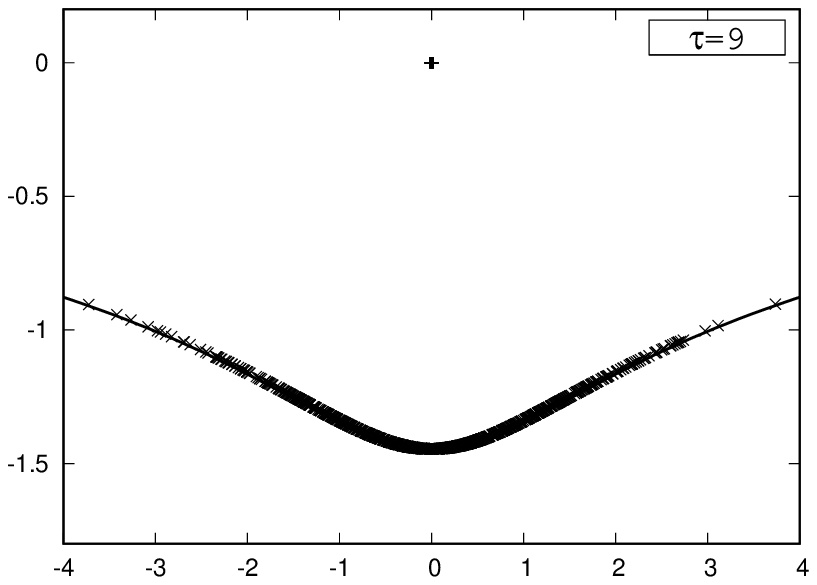}
\caption{The distribution of $z$ (represented by ``$+$'') and
flowed configurations (represented by ``$\times$'')
obtained by applying the RLM  to the phase quenched model
(\ref{single variable Z deform-PQ})
are plotted for $\tau=0,3,6,9$ from Top-Left to Bottom-Right.
The solid line represents the Lefschetz thimble.
At $\tau=0$, the distribution of $z$ and 
flowed configurations coincides.}
\label{fig case3}
\end{figure}

We focus on the choice of parameters $p=4$ and $\alpha=4.2$,
for which there exists only one Lefschetz thimble
and the ordinary CLM is known to be justified \cite{Nagata:2016vkn}.
The flow equations 
\eqref{hol hge-phi}, \eqref{hol hge-J} and \eqref{hol hge}
are solved for $\tau=0,3,6,9$
by using the fourth-order Runge-Kutta algorithm 
with the flow time discretized by
$\Delta\sigma=10^{-3},10^{-4},10^{-5}$ for $\tau=3,6,9$, respectively.
%
The Langevin simulations are performed 
by using the higher-order algorithm \cite{Aarts:2011zn}
with the step-size 
$\epsilon=10^{-5}$ for $\tau=0,3,6$ and
$\epsilon=10^{-6}$ for $\tau=9$.
We choose $z=0$ as the initial configuration,
and for thermalization, we discard the first 
$10^5$, $10^4$ and $10^2$ configurations for $\tau=0$, 
$\tau=3,6$ and $\tau=9$, respectively.
Measurements are made with $10^4$ configurations obtained
every $10^5$, $10^3$, $10^2$, $10$ steps for $\tau=0,3,6,9$, 
respectively. 
In all the three cases (i)-(iii),
we have checked that 
the obtained expectation values of $x^2$ and $x^4$ 
agree well with the exact results.
In Figs.~\ref{fig case1}, \ref{fig case2} and \ref{fig case3},
we show the distribution of $z$ and 
flowed configurations obtained 
in the case (i), (ii) and (iii), respectively. 
The zoom up of the distribution of $z$ is presented
in Appendix \ref{sec:appendix} for the cases (i) and (ii).


Let us first discuss the case (i) shown in Fig.~\ref{fig case1}.
At $\tau=0$, the distribution of $\phi(z;\tau=0)=z$ 
is nothing but that of the ordinary CLM.
At $\tau=3$, the distribution of $z$ comes close to the real axis,
which occurs because the sign problem is reduced to large extent
by the holomorphic gradient flow.
As a result of this,
the distribution of $\phi(z;\tau)$ comes close to a curve,
which looks similar to the one obtained in the GLTM for the same $\tau$
shown in Fig.~\ref{fig case3}.
At larger $\tau$, however,
the distribution of $\phi(z;\tau)$ spreads out again,
and its boundary approaches the Lefschetz thimble.
The spreading of the distribution
can be understood as the effects of
the complex Langevin dynamics,
which takes care of the residual sign problem arising 
from the Jacobian even in the large-$\tau$ limit.
The distribution of $z$ shrinks towards the origin,
but the zoom up in Appendix \ref{sec:appendix} shows that
it does not contract to a distribution on the real axis.
It is interesting that
the expectation values evaluated
with flowed configurations $\phi(z;\tau)$ 
reproduce the exact results for any $\tau$ without reweighting
although the distribution of $\phi(z;\tau)$ looks quite different
for different $\tau$.


The case (ii) shown in Fig.~\ref{fig case2} is of particular interest.
At $\tau=0$, 
the distribution of $\phi(z;\tau=0)=z$ 
is nothing but that of the ordinary CLM, as in the case (i).
As $\tau$ increases, 
the distribution of $\phi(z;\tau)$ approaches the Lefschetz thimble,
while the distribution of $z$ approaches 
the real axis (See also Appendix \ref{sec:appendix}.).
This is due to the fact that the sign problem 
that the complex Langevin dynamics has to take care of
vanishes in the large $\tau$ limit
since the phase of the Jacobian is taken care of by reweighting
and the remaining phase factor
is taken care of by the holomorphic gradient flow.
Thus, in this case our combined formulation interpolates
the CLM and the original Lefschetz-thimble method via the flow time.

In the case (iii) shown in Fig.~\ref{fig case3},
our formulation reduces to the GLTM \cite{Alexandru:2015sua}.
At $\tau=0$, the calculation simply amounts to applying 
the reweighting method
to the original model (\ref{single variable Z}).
At $\tau > 0$, the distribution of $\phi(x;\tau)$ 
is restricted to the deformed integration contour $M_\tau$, which 
approaches the Lefschetz thimble as $\tau$ is increased.
Note also that the distribution of $x$ shrinks towards the origin 
as $\tau$ increases.


\section{Summary and discussions} 
\label{sec:summary}

We have proposed a formulation which combines the CLM and the GLTM
by applying the CLM to the real variables 
that parametrize the deformed integration contour 
in the GLTM.
The residual sign problem in the GLTM is treated in three different
ways. 
We have applied the three versions
to a single-variable model in the case with a single Lefschetz thimble,
and investigated the distribution of flowed configurations,
which are used in evaluating the observables.

In the first version,
where we remove the residual sign problem completely
by the complex Langevin dynamics, the distribution
of flowed configurations changes with the flow time,
and it does not contract to a curve in the 
long flow-time limit
although the Lefschetz thimble appears at the edge of the distribution.
This version is interesting from the viewpoint of 
generalizing the CLM in the line 
of Refs.~\cite{Salcedo:1996sa,Salcedo:2007ji,Wosiek:2015bqg,Seiler:2017vwj}
since the distribution
of flowed configurations for different flow time
can reproduce the expectation values of observables
equally well without the residual sign problem.
In the second version,
where we quench the phase
of the Jacobian and treat it by reweighting,
the distribution of flowed configurations 
contracts to the Lefschetz thimble in the long flow-time limit.
Thus, in this case our formulation interpolates the ordinary CLM
and the original Lefschetz-thimble method.
The third version is nothing but the GLTM with the RLM used
as the Monte Carlo method for updating the real variables
parametrizing the deformed contour.
The distribution of flowed configurations is restricted to
the deformed contour, which approaches the Lefschetz thimble
in the long flow-time limit.

While our formulation is useful in clarifying the relationship
between the CLM and the GLTM, it is not of much practical use
since the computational cost required in solving
\eqref{hol hge-phi}, \eqref{hol hge-J} and \eqref{hol hge}
increases as $O(V)$, $O(V^3)$ and $O(V^5)$, respectively,
with the system size $V$.
Note also that the computational cost increases linearly 
with the flow time. 
Furthermore, our formulation as it stands does not help
in enlarging the range of 
applicability of the CLM \cite{Aarts:2009uq,Aarts:2011ax,Nagata:2016vkn}.
The reason for this is that the distribution of flowed configurations
is qualitatively not much different from 
the distribution of configurations obtained in the CLM,
which is nothing but the distribution obtained for
$\tau=0$ in the first and second versions of our formulation.
For instance, in the model (\ref{single variable Z})
it is known that the CLM gives wrong results 
for $\alpha \lesssim 3.7$ with $p=4$ 
due to the singular drift problem \cite{Nagata:2016vkn}. 
This cannot be cured by 
the first and second versions of our formulation.
For large $\tau$, the situation gets even worse
since the flowed configurations tend to come closer 
to the singularity of the drift term,
and the simulation itself becomes unstable.
Thus, when the CLM fails, our formulation fails as well,
except for the third version, which does not rely on the justification
of the CLM.

Nevertheless, we consider that the relationship of the two methods as seen
in our combined formulation is useful 
in developing these methods further.
In particular, it is possible to generalize the CLM in such a way
that the range of applicability is enlarged as we report 
in a separate paper \cite{in-prep}.


\section*{Acknowledgements}

The authors would like to thank
G.~Basar, K.~Oda and Y.~Yamamoto for valuable discussions.
J.~N.\ was supported in part by Grant-in-Aid 
for Scientific Research (No.\ 16H03988)
from Japan Society for the Promotion of Science.
S.~S.\ was supported by the MEXT-Supported Program for the Strategic
Research Foundation at 
Private Universities ``Topological Science'' (Grant No.\ S1511006).

\appendix

\section{Zoom up of the distribution of $z$}
\label{sec:appendix}

\begin{figure}[ht]
\centering
\includegraphics[width=7cm]{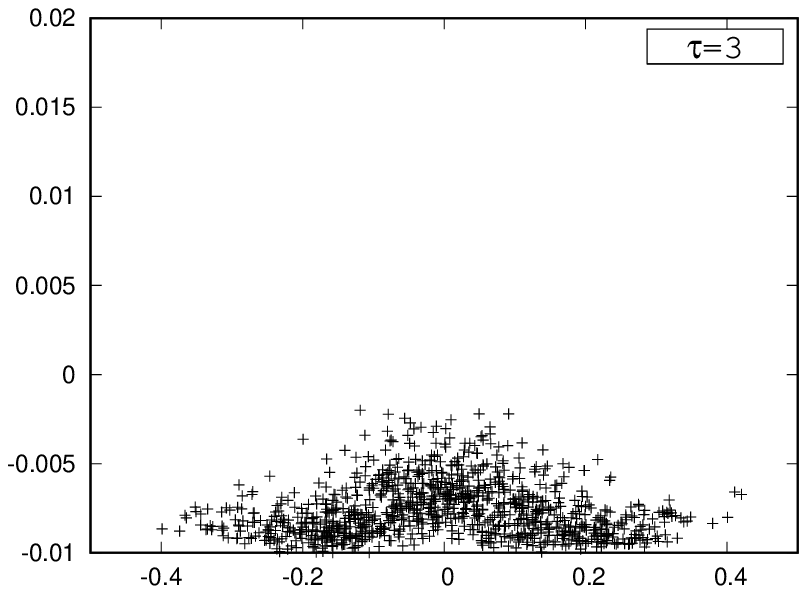}
\includegraphics[width=7cm]{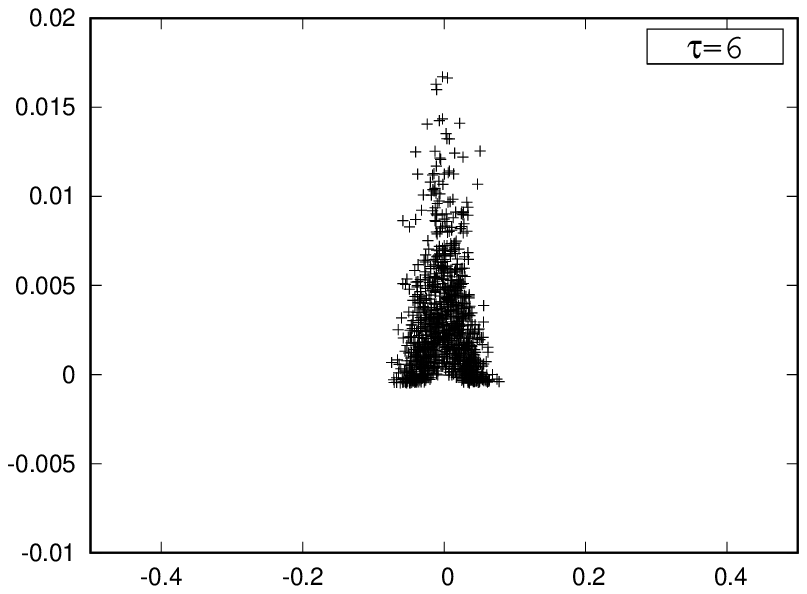}
\includegraphics[width=7cm]{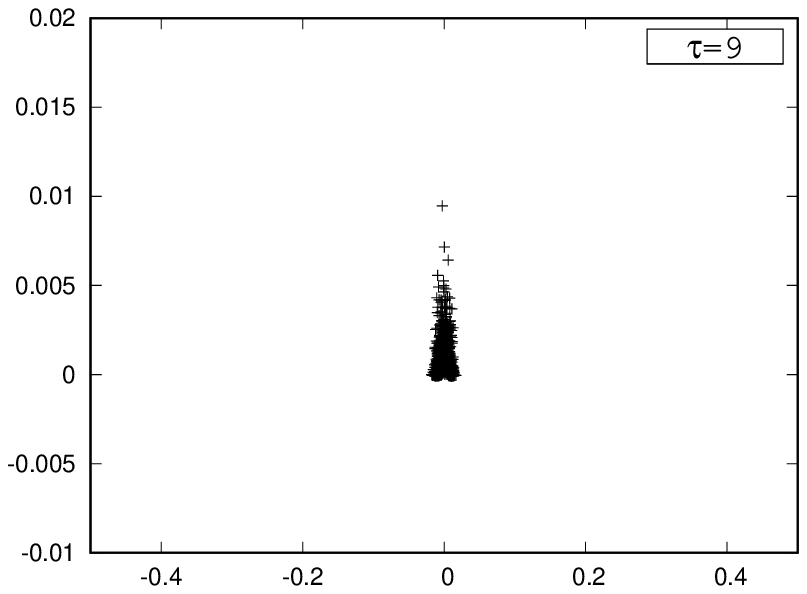}
\caption{Zoom up of the distribution of $z$ in the case (i)
for $\tau=3,6,9$.}
\label{fig case1 zoom up z}
\end{figure}

\begin{figure}[ht]
\centering
\includegraphics[width=7cm]{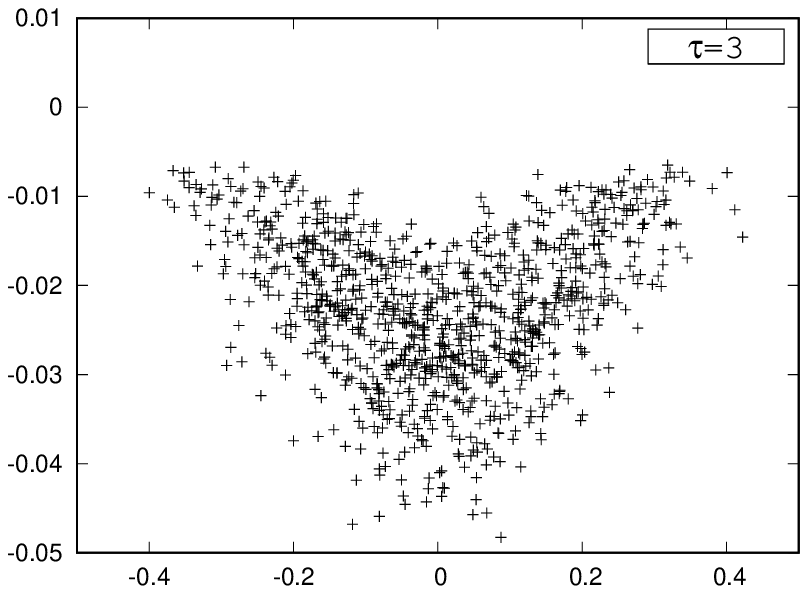}
\includegraphics[width=7cm]{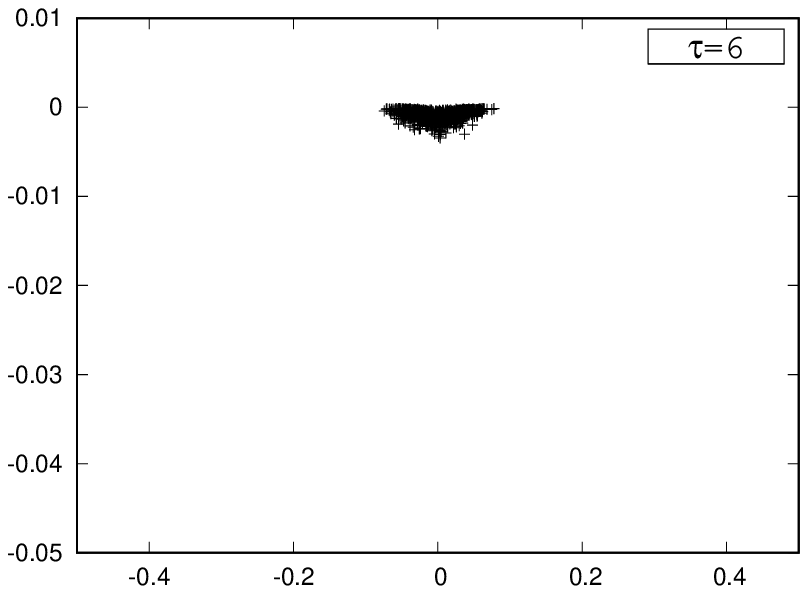}
\includegraphics[width=7cm]{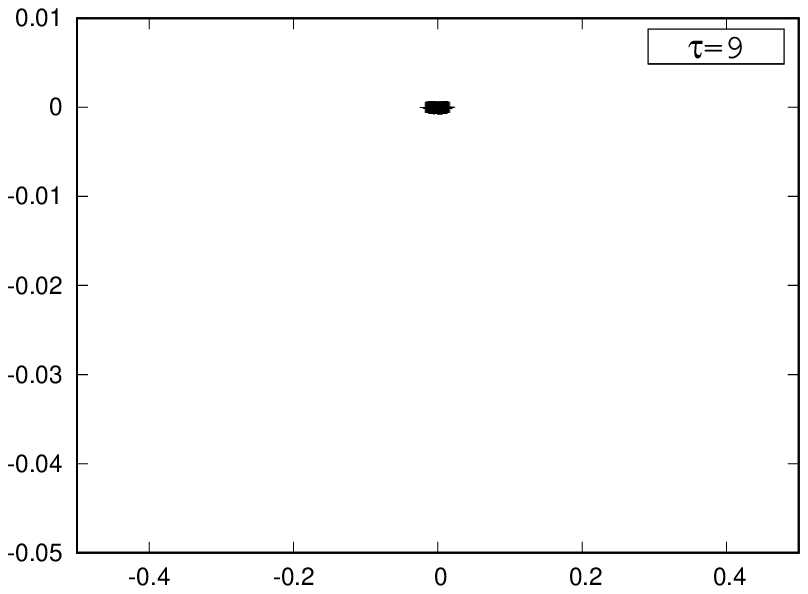}
\caption{Zoom up of the distribution of $z$ in the case (ii)
for $\tau=3,6,9$.}
\label{fig case2 zoom up z}
\end{figure}

In Figs.~\ref{fig case1 zoom up z} and \ref{fig case2 zoom up z},
we present zoom up of
the distribution of $z$ in the case (i) and (ii), respectively,
for $\tau=3,6,9$.
We find that the distribution 
contracts to the real axis in the case (ii),
but not in the case (i).

\end{document}